\documentclass[a4paper,11pt]{article}
\usepackage{pos}

\usepackage{caption}
\usepackage{textcomp}
\usepackage{amsfonts} 
\usepackage{amsmath} 
\usepackage{slashed}
\usepackage{pifont}
\usepackage{multirow,lscape}
\usepackage{array}
\usepackage{subcaption}
\usepackage{verbatim}
\usepackage{bm}
\usepackage{comment}
\usepackage{xcolor}
\usepackage{url} 

\graphicspath{{./fig/}}
\definecolor{jwapurple}{HTML}{7953EE}
\definecolor{uwpurple}{HTML}{4b2e83}
\newcommand{\oper}{\mathcal{O}}

\newcommand{\corr}[1]{\left<{#1}\right>}
\newcommand{\Vcb}{\left|V_{cb}\right|}

\allowdisplaybreaks

\title{Progress report on testing robustness of the Newton method in %
data analysis on 2-point correlation function using a MILC HISQ ensemble}
\ShortTitle{Data analysis on 2pt correlators}

\author[a]{Tanmoy Bhattacharya}
\author[b,1]{Benjamin J. Choi}
\author[a]{Rajan Gupta}
\author[c,1]{Yong-Chull Jang}
\author*[b,1]{Seungyeob Jwa}
\author[b,1]{Sunghee Kim}
\author[b,1]{Sunkyu Lee}
\author[b,1]{Weonjong Lee}
\author[d,1]{Jaehoon Leem}
\author[b,1]{Jeonghwan Pak}
\author[e,1]{Sungwoo Park}

\affiliation[a]{Los Alamos National Laboratory, MS B285, P.O. Box 1663, Los Alamos, NM 87545-0285, USA}
\affiliation[b]{Lattice Gauge Theory Research Center, CTP, and FPRD, 
Department of Physics and Astronomy, Seoul National University, Seoul 08826, South Korea}
\affiliation[c]{Physics Department, Brookhaven National Lab, Upton, NY 11973, USA} 
\affiliation[d]{Computational Science and Engineering Team, Innovation Center, 
Samsung Electronics, Hwaseong, Gyeonggi-do 18448, South Korea}
\affiliation[e]{Lawrence Livermore National Lab, 7000 East Ave, Livermore, CA 94550, USA}
\note{The SWME collaboration}

\emailAdd{wlee@snu.ac.kr}
\emailAdd{thoth@snu.ac.kr}

\abstract{ We report recent progress in data analysis on the two point
  correlation functions which will be prerequisite to obtain
  semileptonic form factors for the $B_{(s)} \to D_{(s)}\ell\nu$
  decays. We use a MILC HISQ ensemble for the measurement.  We use the
  HISQ action for light quarks, and the Oktay-Kronfeld (OK) action for
  the heavy quarks ($b$ and $c$). We used a sequential Bayesian method
  for the data analysis. Here we test the new fitting methodology of
  Benjamin J.~Choi in a completely independent manner. }

\FullConference{%
The 40th International Symposium on Lattice Field Theory,\\
July 31-August 4, 2023,\\
Fermilab, Batavia, Illinois, USA
}

\begin{document}
\maketitle

%
%
%
\section{Introduction}
\label{sec:intr}
Semileptonic form factors for the $B \to D^{(*)}l\nu$ decays can probe
the Cabibbo-Kobayashi-Maskawa (CKM) matrix element $\Vcb$
\cite{HFLAV:2022}.
This requires the precise data analysis on 2-point correlation
functions \cite{Flab:2014}.
Here, we present our recent progress in data analysis on 2-point
correlation functions.
This work provides an independent cross-checking of the methodology
developed by Benjamin Choi in Ref.~\cite{Ben:2021}.
We use the MILC HISQ ensembles for the numerical study.
In Table \ref{tab:intr-1} (\subref{stab:MILC-ens-1}), we summarize
details on the MILC lattice \texttt{a12m310}.
For bottom and charm quarks, we use the Oktay-Kronfeld (OK) action
\cite{OK:2008ex}.
In Table \ref{tab:intr-1} (\subref{stab:hop-1}), we summarize details
on hopping parameters for bottom quarks, and light quark masses.
\begin{table}[h!]
  \begin{subtable}{0.51\linewidth}
    \center
    \resizebox{0.99\linewidth}{!}{
      \begin{tabular}{c|c|c|c}
        \hline\hline
        ID       & a (fm)    & $M_{\pi}$(MeV)&$L^{3}\times T$
        \\ \hline
        \texttt{a12m310}  & 0.1207(11)   & 305.3(4)  &$24^{3}\times 64$
        \\ \hline\hline
      \end{tabular}
    } 
    \caption{Details on the MILC lattice ensemble}
    \label{stab:MILC-ens-1}
  \end{subtable}
  \hfill
  \begin{subtable}{0.47\linewidth}
    \center
    \resizebox{0.99\linewidth}{!}{
      \begin{tabular}{c|c|c|c}
        \hline\hline
        $\kappa_\text{crit}$ & $\kappa_{b}$  & $m_{l}$  & $N_\text{cfg} \times N_\text{src}$
        \\ \hline
        0.051211   & 0.04102  & 0.0509 & $1053 \times 3$
        \\ \hline\hline
      \end{tabular}
    } 
    \caption{Hopping parameters}
    \label{stab:hop-1}
  \end{subtable}
  \caption{ Parameters for the numerical study. Here, $N_\text{cfg}$
    ($N_\text{src}$) represents the number of gauge configurations
    (the number of measurements per gauge configuration).  }
  \label{tab:intr-1}
\end{table}

%

%
%

%
%
%
\section{Fit Function}
Spectral decomposition of 2-pt correlation functions $C(t)$ measured
on the lattice is
\begin{align}
  C(t)=&\sum_{\tau}\corr{\oper^{\dagger}_{\tau}(x)\oper_{\tau}(0)}
  \nonumber \\ 
  =&\sum_{n=0}^{\infty}(-1)^{n(t+1)}
  \left|\corr{n|\oper(0)|0}\right|^{2}
  \left( e^{-E_{n}t}+e^{-E_{n}(T-t)} \right).
\end{align}
where the heavy-light meson operator $\oper$ is
\begin{align}
  \oper_{\tau}(x) &\equiv
  \left[\bar{\psi}(x) \Gamma \Omega(x)\right]_{\tau}\chi(x) \,,
  \\
  \Gamma &= \gamma_{5} \quad \text{or} \quad \gamma_j \,,
  \\
  \Omega(x) &= \gamma_{1}^{m_{1}}\gamma_{2}^{m_{2}}\gamma_{3}^{m_{3}}
  \gamma_{4}^{m_{4}} \,,
  \\
  x_\mu &= m_\mu a \quad \text{with} \quad m_\mu \in Z 
\end{align}
with staggered quark field $\chi(x)$, and the $\psi$ is a heavy quark
field in the OK action \cite{OK:2008ex}.
Here, $n$ is an integer index for energy eigenstates.
If $n$ is even (odd), its eigenstate has even (odd) time-parity.
The oscillating terms with odd time-parity come from the temporal
doubler of staggered quarks \cite{Wingate:2002}.
The $\tau$ represents taste degrees of freedom for staggered quarks.
In our notation, $\lvert 0\rangle$ represents not the vacuum state
($\lvert \Omega \rangle$) but the heavy-light meson ground state with
energy $E_{0} > 0$.

Then, the $n+m$ fitting function \cite{Wingate:2002} is 
\begin{align}
f^{n+m}(t) =  &g^{n+m}(t)+g^{n+m}(T-t), \label{eq:fitfunc}\\
g^{n+m}(t)=&A_{0}e^{-E_{0}t}\Big[1+r_{2}e^{-\Delta E_{2}t}\times\left(1+\dots\times(1+r_{(2n-2)}e^{-\Delta E_{(2n-2)}t})\cdots\right)\nonumber \\
&-(-1)^{t}r_{1}e^{-\Delta E_{1}t}\big(1+\dots\times(1+r_{(2m-1)}e^{-\Delta E_{(2m-1)}t})\cdots\big)\Big]
\end{align}
where $r_{i}=\displaystyle\frac{A_{i}}{A_{i-2}}$, $\Delta
E_{i}=E_{i}-E_{i-2}$ for $i \geq 1$ with $E_{-1} = E_0$ and $A_{-1} =
A_0$.
The $n+m$ fit implies that we include the $n$ even time-parity states
and $m$ odd time-parity states in the fitting function.

%

%
%
%
\section{Newton method}
When we do the least $\chi^{2}$ fitting, we use the
Broyden-Fletcher-Goldfarb-Shanno (BFGS) algorithm \cite{ BBFGS:1970,
  FBFGS:1970, GBFGS:1970, SBFGS:1970} to minimize $\chi^{2}$.
This algorithm belongs to a category of the quasi-Newton method in an
optimization problem \cite{Fletcher:2000}.
The quasi-Newton method requires an initial guess as input to the
fitting by construction.
%

If a initial guess is good, the quasi-Newton method finds the minimum
efficiently.
If the initial guess is poor (out of radius of convergence), the
number of iterations increases and the chance to find the local minima
instead of the global minimum grows up.
%
%
Therefore, it is essential to find a good initial guess close to the
true solution (the global minimum), if we want to save computing
resources.
For this purpose, it is best to obtain initial guess directly from the
data, as long as the computing time is negligibly small compared with
that of the $\chi^2$ minimizer.
It is the multi-dimensional Newton method combined with a scanning
method \cite{Ben:2021} that satisfies these conditions.

%
%
The (1-dimensional) Newton method converges quadratically with respect
to the distance from the true solution.
Hence, it find a root in a few iterations, if the initial guess
is sufficiently close to the true solution.
However, if the initial guess is sufficiently far away from the true
solution, it loses its merit completely.
The scanning method converges slowly, but it can narrow down a range
to find roots in a few iterations.
If we combine the Newton method with the scanning method, it is
possible to keep only the merits, while discarding the disadvantages.
In other words, the scanning method finds a narrow range to find roots
in a few iterations, and then the Newton method can find a root in a
few iterations within the narrow range.
Let us consider the $1+1$ fit as an example.
We feed results for the $1+0$ fit as an initial guess for
the $A_0, E_0$ part of the $1+1$ fit parameters.
In order to obtain initial guess for the remaining parameters $r_1$
and $\Delta E_{1}$, we use the 4-dimensional Newton method combined
with the scanning method.
The initial range is set to $r_{1}\in [0, 1.5]$ and $\Delta E_{1}\in
[0, 1.0]$ for the scanning method.
After a few iterations ($1 \sim 2$), the scanning method finds such a
narrow range for $r_1$ and $\Delta E_1$ that we may use the Newton
method \cite{ Ben:2021} to find an exact solution for Eqs.~\ref{eq:qt}
in a few iterations ($7 \sim 10$).
The Newton method uses the same time slice combination (\emph{e.g.} $
\{ t_1=t_{\min}, t_2, t_3, t_4\} $ for the $1+1$ fit) as the scanning
method.
First, we collect all the possible time slice combinations within the
fit range ( $t_{\min} \le t \le t_{\max}$\,; \emph{e.g.} $t_{\min} = 12$
and $t_{\max} = 26$ for the $1+1$ fit), before using the scanning and
Newton methods.
A possible time slice combination should satisfy the following two
conditions: 1) it must contains $t_{\min}$, and 2) the number of even
time slices must be equal to that of odd time slices.
For each time slice combination, we run the Newton method until we consume
all the time slice combinations.
If the Newton method find a good solution in a few iterations ($7 \sim
10$), then we keep it, and if it fails, we discard that time slice
combination.
The failure rate is about $70\%$ for the 1+1 fit.
The Newton method finds a solution to satisfy the $N$ equations:
$q(t_i)=0$ with $t_i \in \{t_{\min},t_{2},t_{3},\ldots,t_{N}\}$, with
$N=2(n+m)$ for the $n+m$ fit (\emph{e.g.} $N=4$ for the $1+1$ fit).
\begin{equation}
  q(t_{i})=\frac{f(t_{i})-C(t_{i})}{C(t_{i})} = 0\,.
  \label{eq:qt}
\end{equation}
Here, $C(t_{i})$ is the 2pt correlator data coming from our numerical
measurements, and $f(t)$ is the fitting function in
Eq.~(\ref{eq:fitfunc}).
The stopping condition is
\begin{equation}
\max_{i=1,\dots,N}\left|q(t_{i})\right|<10^{-12}\,.
\end{equation}
%

%
%

%
%
%
\section{Fitting results}
\label{sec:fit-res-1}
Here we describe our sequential Bayesian fitting procedure such as
1+0 fit $\to$ 1+1 fit $\to$ 2+1 fit $\to$ 2+2 fit in detail.
We also present results for the $n+m$ fit in each stage of the
sequential Bayesian method (SBM).
We also explain how to perform stability tests at each stage of the
SBM.

%
%
\subsection{1+0 Fit}
\begin{figure}[t!]
  \center
  \vspace*{-7mm}
  \includegraphics[width=0.70\linewidth]{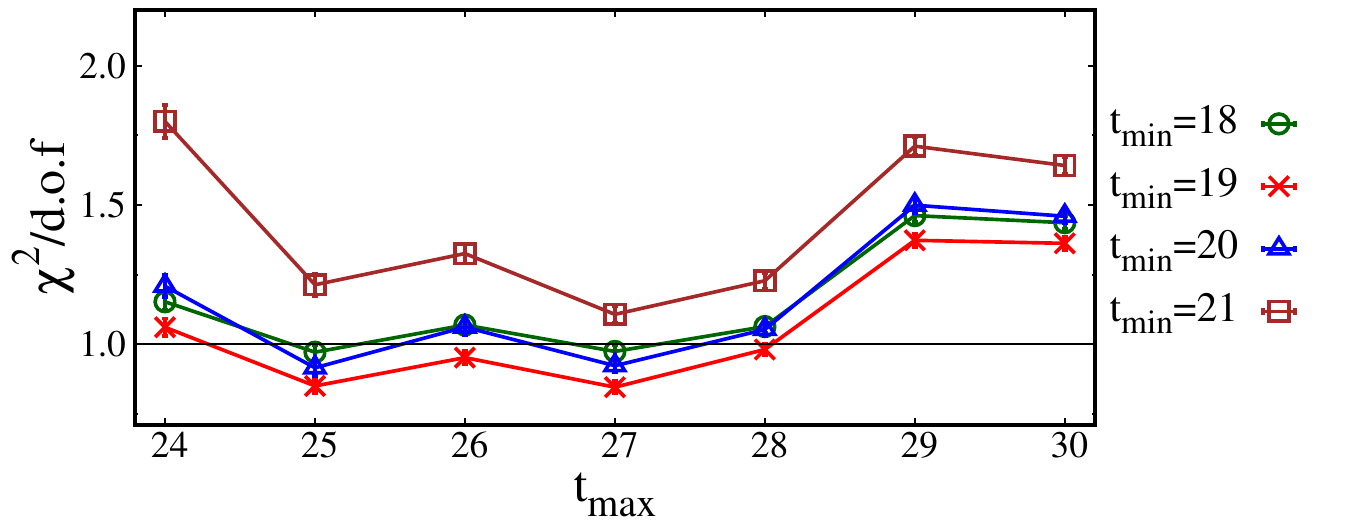}
  \caption{$\chi^{2}/\text{d.o.f}$ of the $1+0$ fit as a function of
    $t_{\min}$ and $t_{\max}$}
  \label{fig:chi_1+0}
\end{figure}

Here the fitting function is $f(t) = f^{1+0}(t)$ in
Eq.~(\ref{eq:fitfunc}).
First we do the $\chi^{2}$ fit for the $1+0$ fit over the whole
appropriate fit ranges.
To find initial guess for $\chi^{2}$ minimizer, we solve the 
following linear equation:
\begin{equation}\label{eq:ig}
\begin{bmatrix}
\sum\limits_{i} \dfrac{C^{2}(t_{i})}{\sigma^{2}(t_{i})} & 
\sum\limits_{i} t_{i} \dfrac{C^{2}(t_{i})}{\sigma^{2}(t_{i})} \\
\sum\limits_{i} t_{i} \dfrac{C^{2}(t_{i})}{\sigma^{2}(t_{i})} &
\sum\limits_{i} t_{i}^{2} \dfrac{C^{2}(t_{i})}{\sigma^{2}(t_{i})} 
\end{bmatrix}
\begin{bmatrix}
\ln A_{0}^{\text{ig}} \\
-E_{0}^{\text{ig}}
\end{bmatrix}
=
\begin{bmatrix}
\sum\limits_{i} \dfrac{C^{2}(t_{i})}{\sigma^{2}(t_{i})}\ln |C(t_{i})| \\ 
\sum\limits_{i} t_{i} \dfrac{C^{2}(t_{i})}{\sigma^{2}(t_{i})}\ln |C(t_{i})|
\end{bmatrix},
\end{equation}
which assumes the diagonal approximation in the covariance matrix \cite{PhysRevD.49.2616}.
Here $A_{0}^{\text{ig}}$, $E_{0}^{\text{ig}}$ denote the initial
guesses, and $\sigma(t_{i})=\sqrt{V(t_{i},t_{i})}$, where
$V(t_{i},t_{i})$ is the covariance matrix of the data $C(t_i)$.
Here the summation $\sum\limits_i$ is over the fit range: $t_{\min}
\leq t_{i} \leq t_{\max}$.
The optimal fit range for the 1+1 fit is determined by the minimum of
$\chi^2/\text{dof}$.
In Fig.~\ref{fig:chi_1+0} we present results for $\chi^2/\text{dof}$
as a function of $t_{\min}$ and $t_{\max}$.
We find that the optimal fit range is $19 \le t \le 28$: $t_{\min} =
19$ and $t_{\max} = 28$.

Note that $t_{\max} = 28$ is fixed for the remaining fits of the SBM.

\begin{figure}[t!]
  \begin{subfigure}{0.61\linewidth}
    \vspace*{-7mm}
    \includegraphics[width=\linewidth]{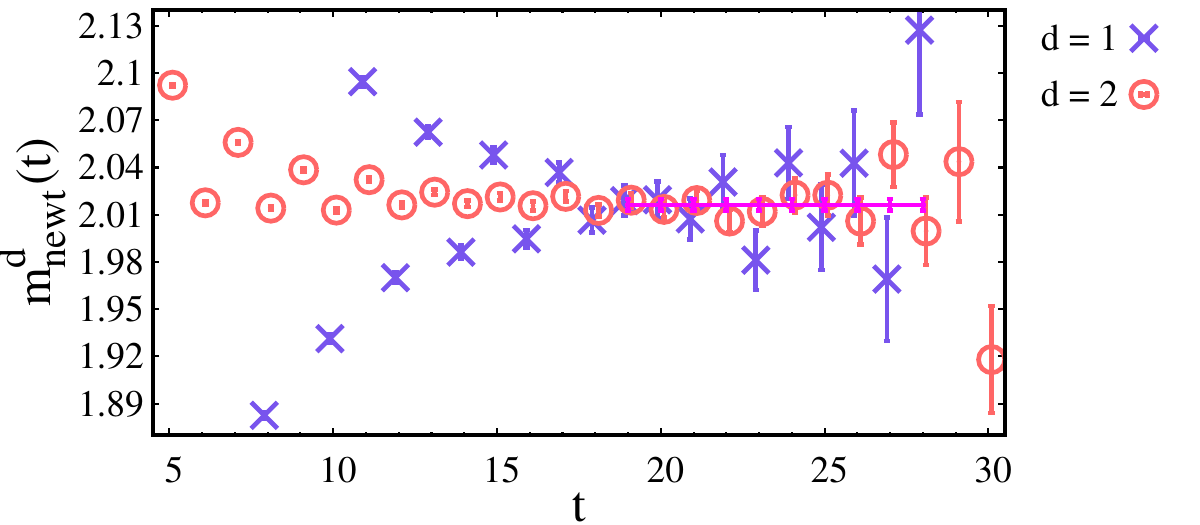}
    \caption{The newton mass plot}
    \label{fig:NM}
  \end{subfigure}
  %
  \hspace{-7mm}
  \begin{subfigure}{0.42\linewidth}
    \vspace*{-7mm}
    \includegraphics[width=\linewidth]{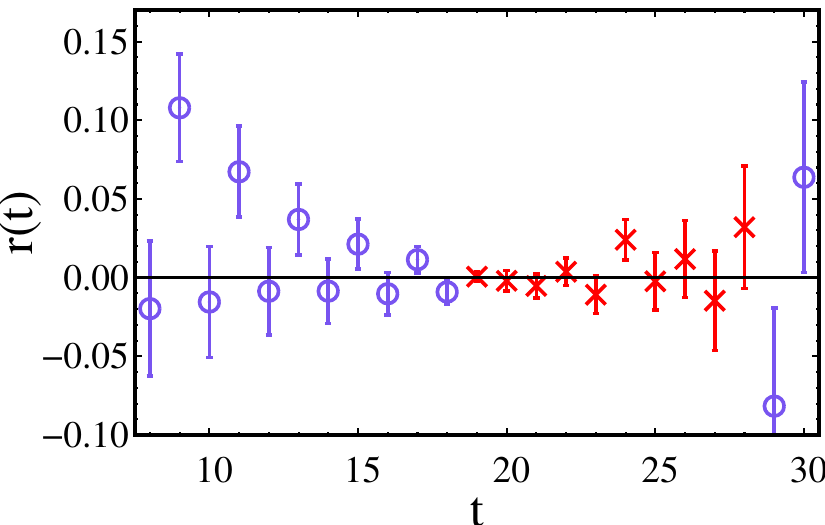}
    \caption{Residual plot of $1+0$ fit}
    \label{fig:resid_1+0}
  \end{subfigure}
  \caption{Results of $1+0$ fit}
  \label{fig:fit_1+0}
\end{figure}

Let us define the Newton mass $m^d_\text{newt}(t) = E_0(t)$ at zero
momentum projection.
At each time slice $t$, we obtain $m^d_\text{newt}(t)$ by solving two
equations: $q(t) = q(t+d) = 0$, using the 2-dimensional Newton
method.
We use the least $\chi^2$ fitting results for $A_0$ and $E_0$ as
initial guess for the 2-d.~Newton method.
In Fig.~(\ref{fig:NM}), we present results for $m^d_\text{newt}(t)$
with $d=1$ and $d=2$ as a function of time $t$.

Results for the 1+0 fit are summarized in Table~(\ref{tab:1+0}).
In order to check the quality of fitting, we plot the residual,
$r(t)=\dfrac{C(t)-f(t)}{\lvert C(t)\rvert}$ in Fig.~\ref{fig:fit_1+0}
(\subref{fig:resid_1+0}).
Here the red (blue) symbols represent the residual within (out of)
the fit range.
\begin{table}[h]
\center
\resizebox{0.65\columnwidth}{!}{
\begin{tabular}{c|c|c|c|c}\hline\hline
Fit range       & $A_{0}$     & $E_{0}$    & $\chi^2/d.o.f.$ & $p$-value \\\hline\hline
[19, 28]        & 0.01735(125)& 2.0162(38) & 0.981(22)       & 0.448(17) \\\hline \hline
\end{tabular}}
\caption{Result of the $1+0$ fit}
\label{tab:1+0}
\end{table}

\subsection{1+1 Fit}
We use results of the 1+0 fit to determine Bayesian priors for the 1+1
fit.
We set the Bayesian prior widths (BPW) to the maximum fluctuation
($\sigma_p^{mf}$) or the signal cut ($\sigma_p^{sc}$) of the $A_0$ and
$E_0$ parameters.\footnote{The signal cut means that the error (=
noise) is the same as the average (= signal), when the parameters
should be positive thanks to physical reasons. }
We choose $\min( \sigma_p^{mf}, \sigma_p^{sc})$ for the BPW.

Starting from $t_{\min}^{1+1} = t_{\min}^{1+0}-2$, we run
$t_{\min}^{1+1}$ over the lower values until $\chi^2/\text{dof}$
overflows the appropriate criterion.
\begin{figure}[t!]
  \begin{subfigure}{0.45\linewidth}
    \includegraphics[width=\linewidth]
                    {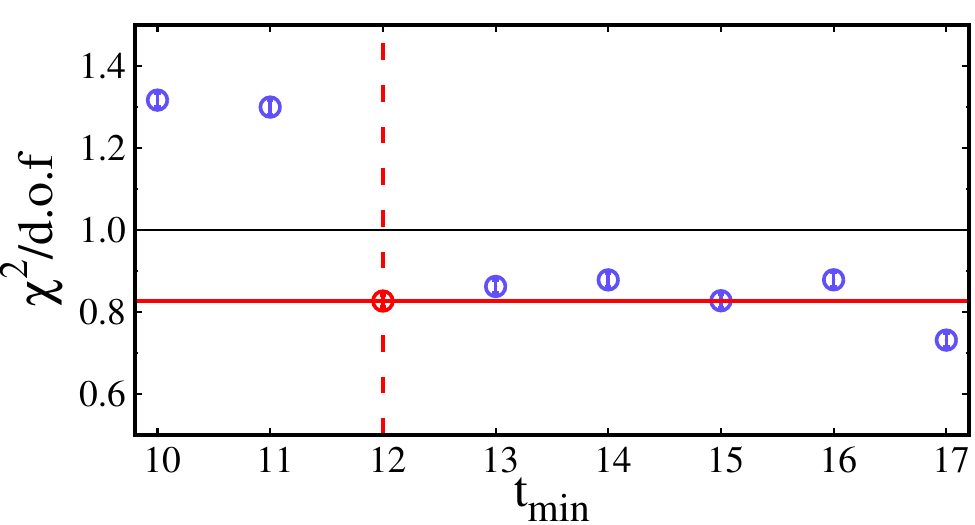}
    \caption{Result of $\chi^{2}/\text{d.o.f}$ of $1+1$ fit}
    \label{fig:chi_1+1}
  \end{subfigure}
  %
  \hspace{1mm}
  \begin{subfigure}{0.540\linewidth}
    \includegraphics[width=\linewidth]
                    {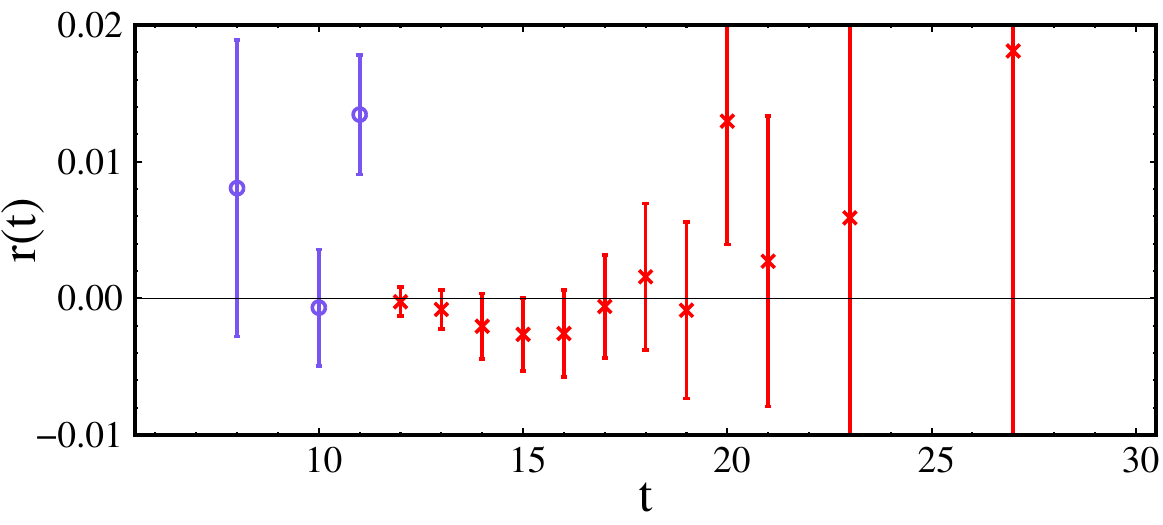}
    \caption{Residual plot of $1+1$ fit}
    \label{fig:resid_1+1}
  \end{subfigure}
  \caption{Results of $1+1$ fit}
  \label{fig:fit_1+1}
\end{figure}

In Fig.~\ref{fig:fit_1+1} (\subref{fig:chi_1+1}) we present
$\chi^2/\text{dof}$ as a function of $t_{\min}^{1+1}$.
Here we find that the optimal fit range is $t_{\min}^{1+1} = 12$.
In Fig.~\ref{fig:fit_1+1} (\subref{fig:resid_1+1}) we present
the residual $r(t)$ with the fit range $12 \le t \le 28$.
The 1+1 fit results are summarized in Table \ref{tab:1+1}.
\begin{table}[b]
\center
\resizebox{0.75\textwidth}{!}{
\begin{tabular}{c |llllll}
\hline\hline
$t_{\min}$ &$A_{0}$ $(10^{-2})$ & $E_{0}$  &$r_{1}$ & $\Delta E_{1}$&$\frac{\chi^2}{\text{d.o.f.}}$\\ \hline
prior &1.735(1735) &2.0162(538)   &     &     &        &  \\
12    &1.847(31)   & 2.0198(13)   & 0.28(14)  & 0.193(39)  & 0.827(15)  \\
\hline\hline
\end{tabular}}
\caption{Result of the $1+1$ fit}
\label{tab:1+1}
\end{table}

%

%
\subsection{2+1 Fit}
\begin{figure}[t!]
  \begin{subfigure}{0.45\linewidth}
    \vspace*{-7mm}
    \includegraphics[width=\linewidth]
                    {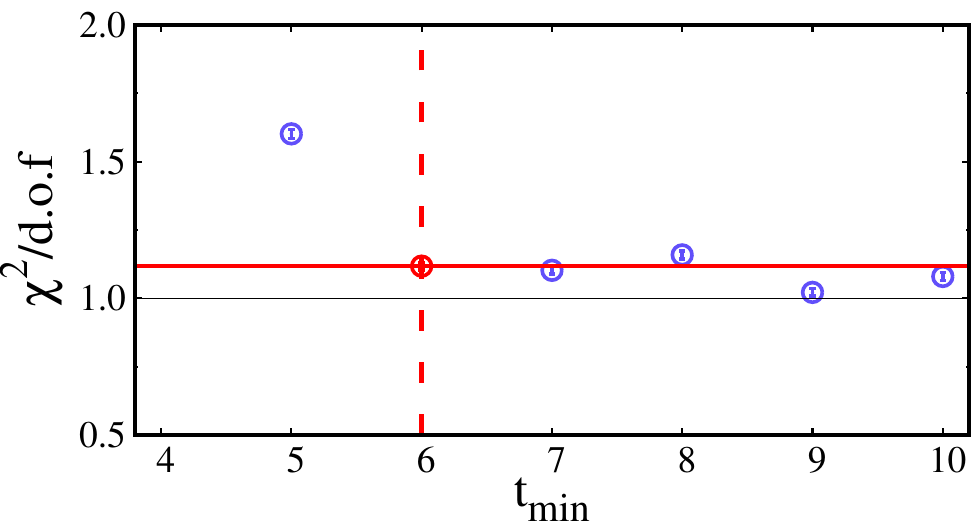}
    \caption{Result of $\chi^{2}/\text{d.o.f}$ of $2+1$ fit}
    \label{fig:chi_2+1}
  \end{subfigure}
  %
  \hspace{1mm}
  \begin{subfigure}{0.540\linewidth}
    \vspace*{-7mm}
    \includegraphics[width=\linewidth]
                    {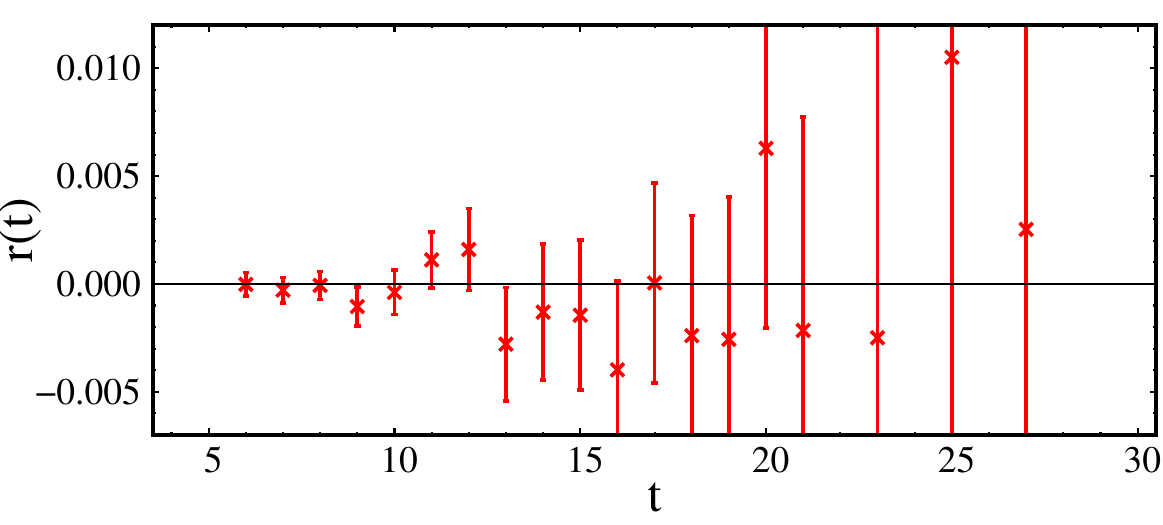}
    \caption{Residual plot of $2+1$ fit}
    \label{fig:resid_2+1}
  \end{subfigure}
  \caption{Results of $2+1$ fit}
  \label{fig:fit_2+1}
\end{figure}

First, we use the results of $1+1$ to set the BPW for $2+1$
fit.
Starting from $t_{\min}^{2+1} = t_{\min}^{1+1}-2$, we run
$t_{\min}^{2+1}$ over the lower values until $\chi^2/\text{dof}$
overflows the criterion.
In Fig.~\ref{fig:fit_2+1} (\subref{fig:chi_2+1}) we present
$\chi^2/\text{dof}$ as a function of $t_{\min}^{2+1}$.
Here we find that $t_{\min}^{2+1} = 6$ is the optimal fit range
for the 2+1 fit.
In Fig.~\ref{fig:fit_2+1} (\subref{fig:resid_2+1}) we present
the residual $r(t)$ with the fit range $6 \le t \le 28$.

\begin{figure}[t!]
  \center
  \begin{subfigure}{0.8\linewidth}
    \vspace*{-7mm}
    \includegraphics[width=\linewidth]
                    {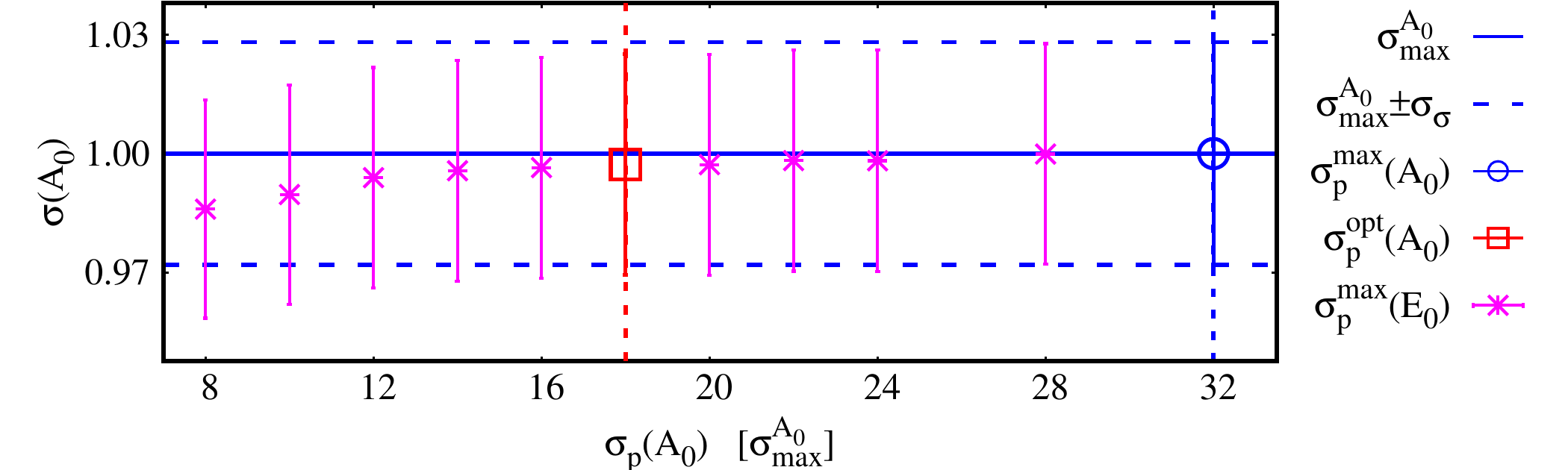}
    \caption{Stability test on $A_{0}$.}
    \label{fig:stab_2+1_A}
  \end{subfigure}
  %
  %
  \begin{subfigure}{0.8\linewidth}
    \includegraphics[width=\linewidth]
                    {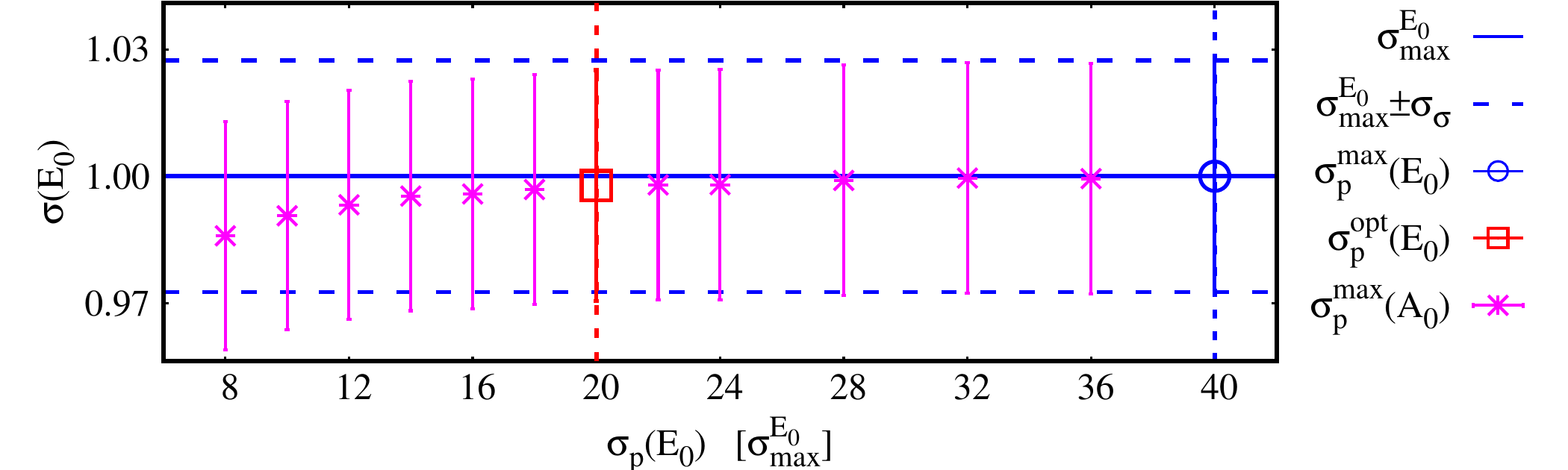}
    \caption{Stability test on $E_{0}$.}
    \label{fig:stab_2+1_E}
  \end{subfigure}
  \caption{Results of the stability test for the $2+1$ fit.}
  \label{fig:stab_2+1}
\end{figure}

Once we choose the fit range, we can perform the stability tests to
find the optimal prior widths for $A_{0}$ and $E_{0}$, which minimize
prior widths with no change in fit results.
Here we adopt the same notation and convention as in Ref.~\cite{Ben:2023}. 
First, we first do the fit with maximum prior widths:
$\sigma_{\text{p}}^{\max} (A_{0}) = \sigma_p^\text{sc}$, and
$\sigma_{\text{p}}^{\max} (E_{0}) = \sigma_p^\text{mf}$, which
correspond to the blue circles in Fig.~\ref{fig:stab_2+1}.
Here note that $\sigma^{A_{0}}_{\max} = \sigma(A_0;
\sigma_p^{\max}(A_0), \sigma_p^{\max}(E_0))$ and
$\sigma^{E_{0}}_{\max} = \sigma(E_0; \sigma_p^{\max}(A_0),
\sigma_p^{\max}(E_0))$.
The units for $x-$axis are $[\sigma^{A_{0}}_{\max}] =
5.54\times10^{-4}$, and $[\sigma^{E_{0}}_{\max}] = 2.01\times10^{-3}$.
%
%
We find the optimal prior widths (the red square symbols,
$\sigma_p^\text{opt} (A_0 \text{ or } E_0)$) such that they are the
minimum prior widths which does not disturb the fit results obtained
with the maximum prior widths.
Here the $\sigma_\sigma$ (blue dashed lines) represents the error of
the error.

Results of the 2+1 fit are summarized in Table \ref{tab:2+1}.
\begin{table}[h]
\center
\resizebox{0.95\textwidth}{!}{
\begin{tabular}{c |lllllll}
\hline\hline
$t_{\min}$ & $A_{0}$ $(10^{-2})$ & $E_{0}$ &  $r_{1}$  & $\Delta E_{1}$ &  $r_{2}$  & $\Delta E_{2}$ &  $\frac{\chi^2}{\text{d.o.f.}}$ \\ \hline
prior & 1.789(997) &2.0180(400) &0.69(69) &0.257(257) & & & \\ \hline
6 &1.789(55) & 2.0180(20) & 0.69(3) & 0.257(6) & 1.04(20) & 0.372(60) & 1.011(14) \\
\hline\hline
\end{tabular}}
\caption{Result of the $2+1$ fit}
\label{tab:2+1}
\end{table}

%


\subsection{2+2 Fit}
\begin{figure}[t!]
  \begin{subfigure}{0.2\linewidth}
    \includegraphics[width=\linewidth]
                    {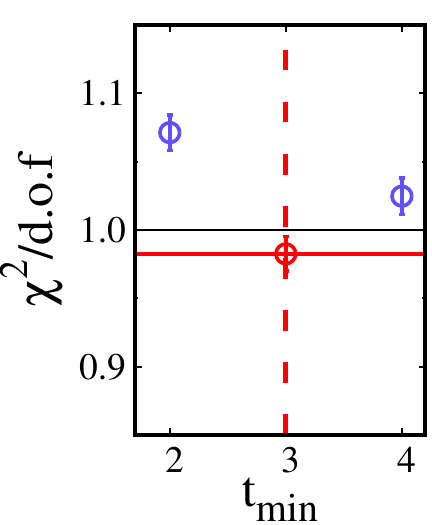}
    \caption{$\chi^{2}/\text{d.o.f}$ of $2+2$ fit}
    \label{fig:chi_2+2}
  \end{subfigure}
	\hspace{1mm}
  \begin{subfigure}{0.398\linewidth}
    \includegraphics[width=\linewidth]
                    {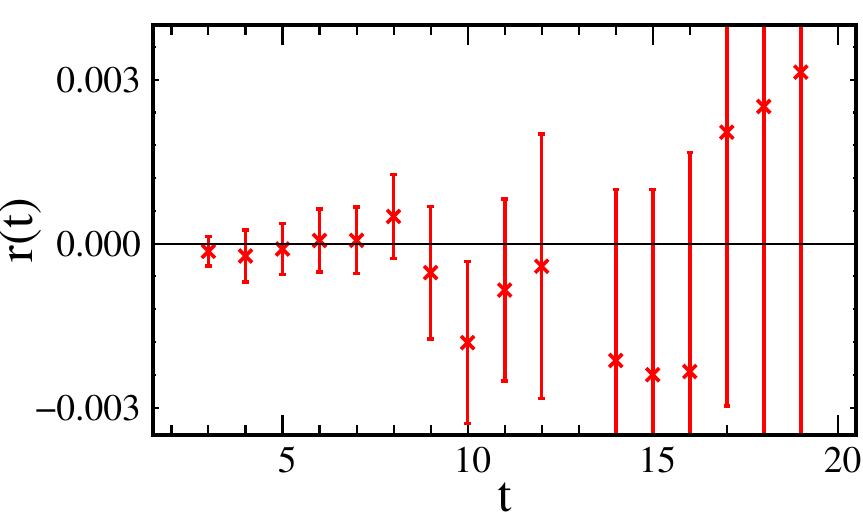}
    \caption{Residual plot of $2+2$ fit (Zoom-in)}
    \label{fig:resid_2+2_in}
  \end{subfigure}
	\hspace{1mm}
  \begin{subfigure}{0.398\linewidth}
    \includegraphics[width=\linewidth]
                    {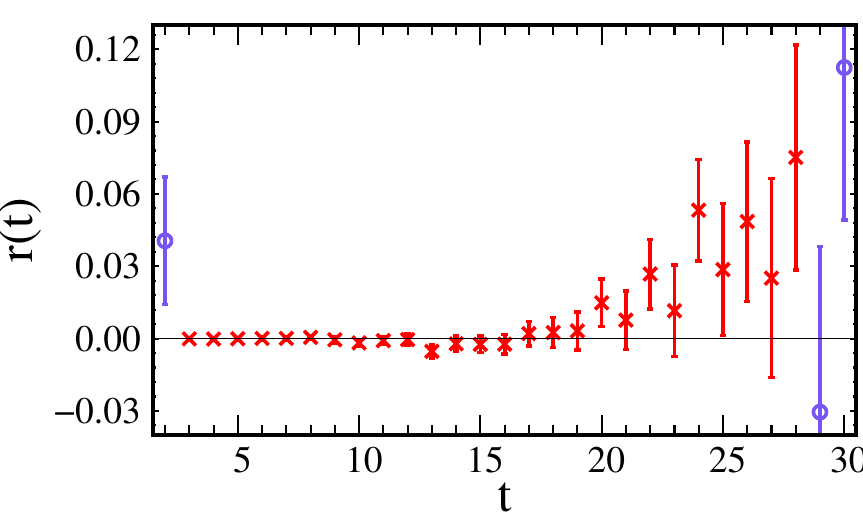}
    \caption{Residual plot of $2+2$ fit}
    \label{fig:resid_2+2_out}
  \end{subfigure}
  \caption{Results of $2+2$ fit}
  \label{fig:fit_2+2}
\end{figure}

For the 2+2 fit we set the prior widths as follows.
\begin{enumerate}
\item {} [$A_0$ and $E_0$] We use the results of the stability tests for
  the 2+1 fit as the prior widths for $A_0$ and $E_0$. 
\item {} [$R_1$ and $\Delta E_1$] We set the prior widths to the
  signal cuts for both.
  %
  %
\item {} [$R_2$ and $\Delta E_2$] We set the prior widths to the
  signal cuts for both.
  %
  %
\item {} [$R_3$ and $\Delta E_3$] No prior information.
\end{enumerate}
Starting from $t_{\min}^{2+2} = t_{\min}^{2+1}-2 = 4$, we run
$t_{\min}^{2+2}$ over the lower values.
In Fig.~\ref{fig:fit_2+2} (\subref{fig:chi_2+2}) we present $\chi^2 /
\text{dof}$ as a function of $t_{\min}^{2+2}$.
The physical positivity \cite{Luscher:1984is} constrains $t_{\min}$ such that
$t_{\min} \ge 3$.
Here we find that $t_{\min}^{2+2} = 3$ is the optimal fit range.
In Fig.~\ref{fig:fit_2+2} (\subref{fig:resid_2+2_out}) we present the 
residual $r(t)$ with the fit range $3 \le t \le 28$.

\begin{figure}[t!]
  \center
  \begin{subfigure}{0.8\linewidth}
    \vspace*{-7mm}
    \includegraphics[width=\linewidth]{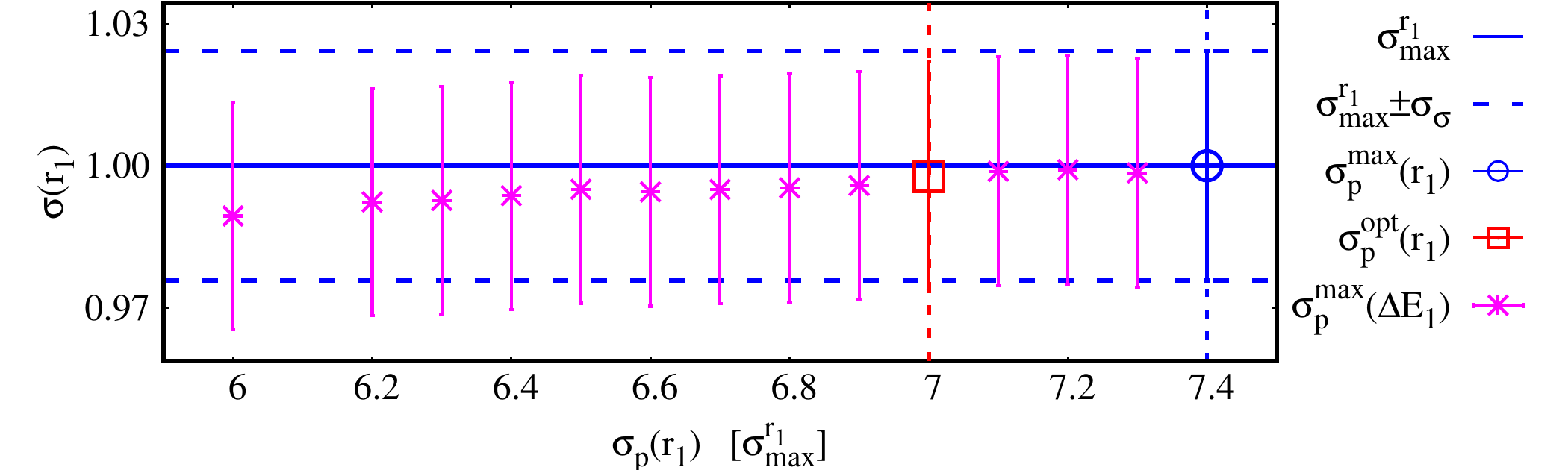}
    \caption{Stability test on $r_{1}$.}
    \label{fig:stab_2+2_r}
  \end{subfigure}
  %
  %
  \begin{subfigure}{0.8\linewidth}
    \includegraphics[width=\linewidth]{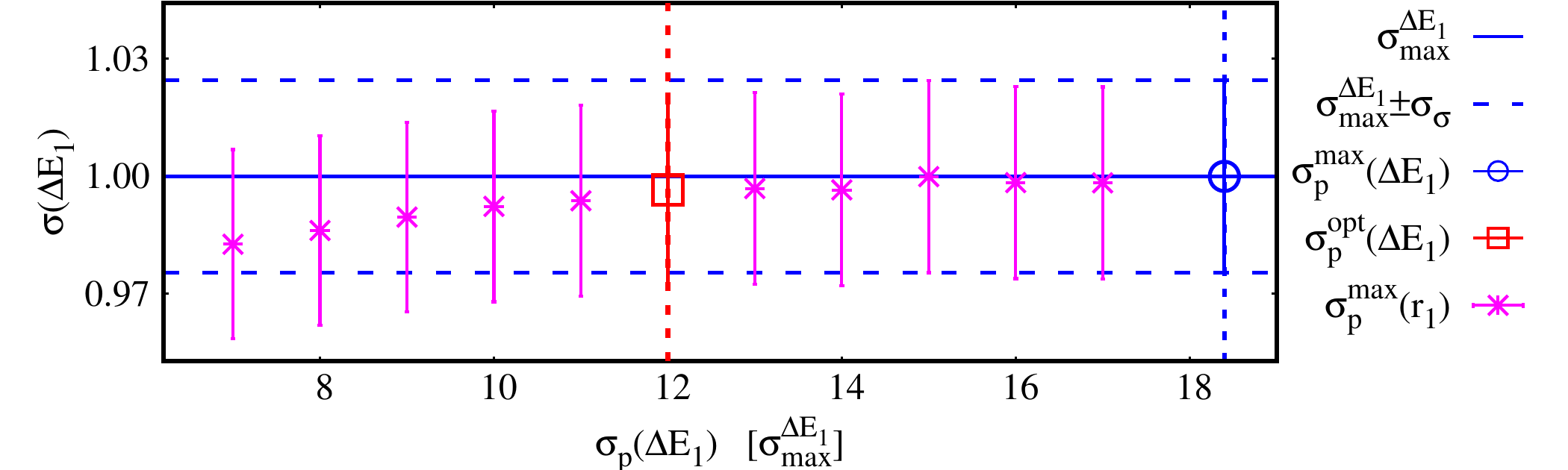}
    \caption{Stability test on $\Delta E_{1}$.}
    \label{fig:stab_2+2_E}
  \end{subfigure}
  \caption{Results for the stability test in the $2+2$ fit.}
  \label{fig:stab_2+2}
\end{figure}

Once we choose the fit range, we do the stability tests to find the
optimal prior widths for $r_{1}$ and $\Delta E_{1}$.
First, we first do the fit with maximum prior widths:
$\sigma_{\text{p}}^{\max} (r_{1})$, and $\sigma_{\text{p}}^{\max}
(\Delta E_{1})$, which correspond to the blue circles in
Fig.~\ref{fig:stab_2+2}.
Here note that $\sigma^{r_{1}}_{\max} = \sigma(r_1;
\sigma_p^{\max}(r_1), \sigma_p^{\max}(\Delta E_1))$ and
$\sigma^{\Delta E_{1}}_{\max} = \sigma(\Delta E_1;
\sigma_p^{\max}(r_1), \sigma_p^{\max} (\Delta E_1))$, while
$\sigma_p^{\max} (r_1) = \sigma_p^\text{sc} (r_1)$ and
$\sigma_p^{\max} (\Delta E_1) = \sigma_p^\text{sc} (\Delta E_1)$.
The units for $x-$axis are $[\sigma^{r_{1}}_{\max}] =
7.8\times10^{-2}$, and $[\sigma^{\Delta E_{1}}_{\max}] =
1.31\times10^{-2}$.
In Fig.~\ref{fig:stab_2+2} the red square symbols represent the
optimal prior widths, $\sigma_p^\text{opt} (r_1 \text{ or } \Delta
E_1)$).

Results of the 2+2 fit are summarized in Table \ref{tab:2+2}.
\begin{table}[h]
\center
\resizebox{0.7\textwidth}{!}{
\begin{tabular}{c|lllllllll}
\hline\hline
$t_{\min}$	& $A_{0}$ $(10^{-2})$	& $E_{0}$			& $r_{1}$	& $\Delta E_{1}$	&											\\ \hline
prior			& 1.858(997)				& 2.0203(400)		& 0.58(53)	& 0.242(171)		&											\\ \hline
3				& 1.858(24)					& 2.0203(11)		& 0.58(8)	& 0.242(13)			&											\\ \hline\hline
				& $r_{2}$					& $\Delta E_{2}$	& $r_{3}$   &$\Delta E_{3}$	& $\frac{\chi^2}{\text{d.o.f.}}$	\\ \hline
prior			& 1.91(191)					& 0.512(512)       &           &						&											\\ \hline
3				& 1.91(6)					& 0.512(17)       & 1.63(19)  &0.480(114)			& 0.983(13)								\\ \hline\hline
\end{tabular}
}
\caption{Result of the $2+2$ fit.}
\label{tab:2+2}
\end{table}

%

%
%

%
\acknowledgments
We would like to thank A.~Kronfeld, and C.~Detar for helpful
discussion on fitting.
We thank the MILC collaboration and Chulwoo Jung for providing the
HISQ lattice ensembles.
The research of W. Lee is supported by the Mid-Career Research Program
(Grant No. NRF-2019R1A2C2085685) of the NRF grant funded by the Korean
government (MSIT).
W. Lee would like to acknowledge the support from the KISTI
supercomputing center through the strategic support program for the
supercomputing application research (No. KSC-2018-CHA-0043,
KSC-2020-CHA-0001, KSC-2023-CHA-0010).
Computations were carried out in part on the DAVID cluster at Seoul
National University.

\bibliography{ref}

\end{document}